\documentclass[10pt]{article}
\usepackage{amsmath, amsfonts, amsthm, amssymb,verbatim}
        \newtheorem{theorem}{Theorem}[section]
\newtheorem{definition}[theorem]{Definition}

        \newtheorem{remark}[theorem]{Remark}

\numberwithin{equation}{section}

\newcommand \Riem       {\text{Rm}} 
\newcommand \inj        {\text {Inj}}
\newcommand \vol     	{\text {Vol}}
\newcommand \Bcal   		{\mathcal B} 
\newcommand \expo   		{\text{exp}}

\newcommand \xbf {{\mbox{\boldmath$x$}}}

\newcommand \Mbf {\mathbf M}
\newcommand \pbf {\mathbf p}

\newcommand \qbf {\mathbf q}

\newcommand \Nbf {\mathbf N}

\newcommand \la \langle
\newcommand \ra \rangle
\newcommand \tbar {{\overline t}}

\newcommand \bart {\underline t}

\newcommand \barc {\underline c}
\newcommand \cbar {\overline c}
\newcommand \Acal {\mathcal A}

\newcommand \bark {\underline k}
\newcommand \Rmax {R_{\text{max}}}

\newcommand \gbf        {\mathbf g}

\newcommand \Tbf        {\mathbf T}

\newcommand \Dbf      {\mbox{\boldmath$\nabla$}}

\newcommand \be     {\begin{equation}}
\newcommand \ee     {\end{equation}}

\newcommand \kbar   {\overline k}
\newcommand \del        \partial
\newcommand \eps     \epsilon
\newcommand \auth   \textsc

\newcommand \Lie    {{\mathcal L}}
\newcommand \Mcal    {{\mathcal M}}

\newcommand \Jcal   {{\mathcal J}}
\newcommand \Ncal   {{\mathcal N}}
\newcommand \Hcal   {{\mathcal H}}

\newcommand \Ical   {{\mathcal I}}

\newcommand \NullInj   {\text{Null Inj}}

\newcommand \Lbf    {\mathbf L}

\begin{document}


\title{Injectivity radius and optimal regularity of Lorentzian manifolds with bounded curvature}
\author{
Philippe G. LeFloch$^1$}

\date{October 3, 2008}

\maketitle

\footnotetext[1]{Laboratoire Jacques-Louis Lions \& Centre National de la Recherche Scientifique,
Universit\'e Pierre et Marie Curie (Paris 6), 4 Place Jussieu,  75252 Paris, France.
\\
E-mail : {pgLeFloch@gmail.com} 
\\
\textit{AMS Subject Classification.} 83C05, 53C50, 53C12.
\textit{Key words and phrases.} Lorentzian geometry, injectivity radius, 
constant mean curvature foliation, harmonic coordinates.
\\
Submitted to : ``Actes du S\'eminaire de Th\'eorie Spectrale et de G\'eom\'etrie''.}

\begin{abstract}
We review recent work on the local geometry and optimal regularity of Lorentzian manifolds
with bounded curvature. Our main results provide an estimate of the injectivity radius 
of an observer, and a local canonical foliations by CMC (Constant Mean Curvature) hypersurfaces,
together with spatially harmonic coordinates. 
In contrast with earlier results based on a global bound for derivatives of the curvature,  
our method requires only a sup-norm bound on the curvature near the given observer.  
\end{abstract}


\section{Introduction}
\label{INT}

In this survey, we investigate a few questions about the 
local geometry and regularity of {\sl pointed Lorentzian manifolds} --in which, by definition,
 a point and a future-oriented, unit time-like vector (an {\sl observer}) have been selected.
We are especially interested in manifolds satisfying Einstein equations of general relativity,
referred to as {\sl spacetimes.}  Our main assumption will be purely geometric, viz.~an a~priori bound 
on the 
curvature of the manifold. In the existing literature, conditions involving the derivatives of the curvature 
are assumed. 

Our purpose is, first,  
to derive an estimate on the injectivity radius of a given observer and, second, to construct local coordinate
charts in which the metric coefficients have the best possible regularity.  
This survey is based on the papers \cite{CL1,CL2} written in collaboration with B.-L. Chen. 
We will present the main statements together with a sketch of the proofs. 
For additional background and details, the reader should refer to \cite{CL1,CL2} and the references therein. 

In Riemannian geometry, it is well-known that normal coordinates (defined from the exponential map) 
fail to achieve the optimal regularity of the metric coefficients. Instead, 
the use of {\sl harmonic coordinates} was 
advocated by De~Turck and Kazdan \cite{DeTurckKazdan}, 
while a quantitative estimate on the {\sl harmonic radius}
(involving curvature and volume bounds, only) 
was later derived by Jost and Karcher \cite{JostKarcher}; see Section~\ref{RIEM}, below.  

A Lorentzian metric, by definition, is not positive definite and 
we need to introduce Lorentzian notions of injectivity radius and curvature bound, 
since standard definitions from Riemannian geometry do not apply.   
As it turns out, it is necessary to fix an observer $(\pbf,\Tbf_\pbf)$ which, 
in a canonical way,
induces a positive-definite, inner product $\gbf_{\Tbf_\pbf}$ on the
tangent space at $\pbf$. By parallel transporting the given vector (using the Lorentzian structure)
to a neighborhood of $\pbf$ we construct a (possibly multi-valued) vector field 
$\Tbf$ and, in turn, a ``reference'' Riemaniannian metric $\gbf_\Tbf$.  

Our main estimate of the Lorentzian injectivity radius, in Section~\ref{INJECT} below, 
is purely local and does not require to fix in advance a foliation nor, a fortiori, 
a local coordinate chart. To the reference metric $\gbf_\Tbf$ we apply     
classical arguments from Riemannian geometry (involving
geodesics, Jacobi fields, and comparison arguments).  
By observing that geodesics in the (flat) Euclidian and Minkowski spaces coincide, we are able to compare the behavior 
of $\gbf_T$-geodesics and $\gbf$-geodesics and, finally, 
to transpose the Riemannian estimates into 
estimates for the Lorentzian metric $\gbf$.

Section~\ref{REGUL} concerns mainly the  class of vacuum Einstein spacetimes and 
is devoted to a construction of ``canonical'' local coordinates defined near the observer. 
Under curvature and injectivity bounds only, we establish the existence of {\sl local coordinates charts} 
that are defined in balls with definite size and in which the metric coefficients have optimal regularity.
The proof is based on {\sl quantitative estimates} valid locally near the observer: these estimates control, 
on one hand, a canonical foliation by spacelike hypersurfaces with constant mean curvature 
and, on the other hand, the metric coefficients expressed in {\sl spatially harmonic coordinates.} 

The results and techniques in this work should be useful in the context of general relativity
for investigating the long-time behavior of solutions to the Einstein equations.

Recall that the first work on the local regularity of Lorentzian metrics is due to
Anderson who, in the pioneering work \cite{Anderson-regularity},
proposed to use normal coordinates and spatially harmonic coordinates; this approach, however, 
does not yield the optimal regularity. 
Anderson assumed a sup-norm bound (plus other foliation conditions)
and initiated an ambitious program to investigate the long-time evolution for the Einstein equations
in connection with Penrose's cosmic conjecture. 

On the other hand, Klainerman and Rodnianski \cite{KR4,KR5} assume an $L^2$ curvature bound 
(plus other foliation conditions) and currently develop a vast program (the $L^2$ curvature conjecture) 
on the vacuum Einstein equations via harmonic analysis techniques. Their main results 
concern the geometry of null cones in vacuum spacetimes,  
rather than the geometry of the spacetime itself; their proofs rely on {\sl hyperbolic PDE's} techniques 
(including harmonic analysis), while our approach is {\sl elliptic} in nature. 

There exists also an extensive study of (sufficiently regular) spacetimes
admitting {\sl global} foliations by spatially compact hypersurfaces with constant mean curvature; 
see, in particular, Andersson and Moncrief \cite{AM1,AM2} 
who, also, advocate the use of CMC-harmonic coordinates. 
In these works, estimates (based on the so-called Bel-Robinson tensor) for third-order derivatives 
of the metric are involved. 
In contrast, we focus here on the {\sl local} existence of foliations but under the sole assumption
that the curvature is {\sl bounded.}  

\section{The case of Riemannian manifolds}
\label{RIEM}

The following theorem summarizes two classical results due 
to Cheeger, Gromov, and Taylor \cite{CGT} and Jost and Karcher \cite{JostKarcher}, respectively.

\begin{theorem}[The case of Riemannian manifolds]  
\label{RRR} 
Let $K_0, V_0$ be positive constants and let $(M,g,p)$ be a complete, pointed Riemannian $n$-manifold 
with boundary 
such that the unit geodesic ball $\Bcal_g(p,1)$ is compactly included in $M$ and 
the following curvature and volume bounds hold 
$$ 
\| \Riem_g \|_{\Lbf^\infty(\Bcal(p,1))} \leq K_0, \qquad \vol_g(\Bcal_g(p,1)) \geq V_0.  
$$  
Then, for some constant $I_0=I_0(K_0,V_0,n) \in (0,1)$ the following properties hold: 
\begin{enumerate}
\item[(i)] The injectivity radius $i(p)$ at the point $p$ is greater than $I_0$, that is,
 the (restriction of the) exponential map   
$\expo_\pbf : B(0,I_0) \subset T_p M \to \Bcal_g(p,I_0) \subset M$
is a diffeomorphism onto its image.  

\item[(ii)] Given $ \eps >0$, there exist harmonic coordinates which 
cover the closed ball $\overline{\Bcal_g(p,I_0)}$ and satisfy 
$$
\aligned 
&  e^{-\eps} \, g_E \leq g \leq e^\eps \, g_E,  
\\ 
&  
\| g \|_{W^{2,a}(\Bcal_g(p,I_0))} \leq C_{\eps,a}, \qquad a \in [1,\infty), 
\endaligned
$$
where $g_E$ denotes the Euclidian metric in the local coordinates, and 
the constant $C_{\eps,a}>0$ depends solely on $\eps>0$ and $a \in [1,\infty)$.  
\end{enumerate} 
\end{theorem}

The expression $\| g \|_{W^{2,a}(\Bcal_g(p,I_0))}$ denotes the standard 
$W^{2,a}$ Sobolev norm of the metric coefficients in the local coordinates under consideration. 
An important feature of the above theorem is that no assumption
is imposed on the derivatives of the curvature tensor for, otherwise, 
the statement would be much weaker and of limited interest for the applications.


\section{Injectivity radius of pointed Lorentzian manifolds}
\label{INJECT}

Let $(\Mbf,\gbf)$ be a time-oriented, $(n+1)$-dimensional Lorentzian manifold with boundary, and let $\Dbf$ 
be the Levi-Civita connection associated with $\gbf$.  
Rather than a single point $\pbf \in \Mbf$ as was sufficient in the Riemannian case,
 we need to prescribe an {\sl
observer,} that is, a pair $(\pbf,\Tbf_\pbf)$ where $\Tbf_\pbf$ is a {\sl reference vector} in $T_\pbf \Mbf$, 
that is, a future-oriented, unit timelike vector.  
We refer to $(\Mbf,\gbf,\pbf,\Tbf_\pbf)$ as a {\sl pointed Lorentzian manifold.} 

The reference vector induces an inner product $\gbf_{\Tbf_\pbf}$ on the
tangent space $T_\pbf \Mbf$, defined as follows.
Let $e_\alpha$ ($\alpha =0,\ldots,n$) be an orthonormal frame at $\pbf$, 
where $e_0=\Tbf_\pbf$ and the vectors $e_j$ ($j=1, \ldots, n$) are spacelike. 
Denoting by $e^\alpha$ the corresponding dual frame, we see that in the tangent space at $\pbf$, 
the Lorentzian metric reads 
$\gbf  = - e^0 \otimes e^0 + e^1 \otimes e^1 + \ldots + e^n \otimes e^n$ 
so that the reference metric is 
$\gbf_{\Tbf_\pbf} := e^0 \otimes e^0 + e^1 \otimes e^1 + \ldots + e^n \otimes e^n$. 

The reference metric $\gbf_\Tbf$ is needed to compute the norms $|A|_{\gbf_{\Tbf_\pbf}}$ of a
 tensor $A$ at the point $\pbf$. In case $\Tbf$ is a vector field defined in a neighborhood of $\pbf$
 then this construction can be done at each point and yields a {\sl reference Riemannian metric} $\gbf_\Tbf$, 
 which is canonically determined from the given vector field. 

Consider now the exponential map $exp_\pbf$ at the point $\pbf$, which is
defined on the Riemannian ball $B_{\gbf_{\Tbf_\pbf}}(\pbf, r) \subset T_\pbf \Mbf$ for
all sufficiently small $r$, at least.

\begin{definition}
\label{radius2} 
The {\sl injectivity radius} $\inj(\Mbf,\gbf,\pbf,\Tbf_\pbf)$ of an observer $(\pbf,\Tbf_\pbf)$
is the supremum among all radii $r$ such that the exponential map $\expo_\pbf$ is a global diffeomorphism 
from $B_{\Tbf_\pbf}(0,r)$ to $\Bcal_{\Tbf_\pbf}(\pbf,r) := \expo(B_{\Tbf_\pbf}(0,r))$.
\end{definition}

To begin with, we present a result which relies on a given foliation of a domain of the spacetime
 $\Omega = \bigcup_{t \in [-1,1]} \Hcal_t$ containing the point $\pbf \in \Hcal_0$. 
Here, $\Hcal_t$ are spacelike hypersurfaces with future-oriented, unit normal vector
$T^\alpha := - \lambda \, \nabla^\alpha t$ and lapse function $ \lambda >0$.
We always assume that the geodesic ball $ \Bcal_{\Hcal_0}(\pbf,1) \subset \Hcal_0$ 
(determined by the induced reference metric $\gbf_{\Hcal_0}$) is compactly contained in $\Hcal_0$. 
We make the following main assumptions:  
\begin{enumerate}

\item[] $(A1)  \quad   e^{-K_0} \leq  \lambda \leq e^{K_0}$. 

\item[] $(A2)  \quad  \sup_\Omega |\Lie_\Tbf \gbf |_{\gbf_\Tbf} \leq K_0$.  
 
\item[] $(A3)  \quad  \sup_\Omega |\Riem_\gbf |_{\gbf_\Tbf} \leq K_0$.

\item[] $(A4)  \quad  \vol_{\gbf_{\Hcal_0}}(\Bcal_{\gbf_{\Hcal_0}}(\pbf,1))\geq v_0$.  

\end{enumerate}

\begin{theorem}[Injectivity radius estimate for a foliation]
\label{premier}
Given foliation constants $K_0,V_0$ and a dimension $n$, there exists a constant $I_0>0$
such that, for every foliation satisfying the assumptions $(A1)$--$(A4)$
near a base point $\pbf \in \Mbf$, 
the injectivity radius at $\pbf$ is uniform bounded below by $I_0$, that is, 
$$ 
\inj(\Mbf,\gbf,\pbf,\Tbf_\pbf) \geq I_0.
$$
\end{theorem}

\begin{proof} We only indicate the main steps of the proof.   
First of all, 
according to Jost and Karcher \cite{JostKarcher} and in view of the curvature bound (A3) and the volume bound (A4) 
 on the initial hypersurface, 
one can introduce harmonic coordinates $(x^j)$ on the initial hypersurface $\Hcal_0$, only. 

Then, we can transport these coordinates to the whole of $\Omega$ by following the integral curves of 
the vector field $\Tbf$. This generates coordinates $(x^\alpha) = (t, x^j)$, in which the Lorentzian 
and Riemannian metrics read
$\gbf = - \lambda^2 \, dt^2 + g_{ij} \, dx^i dx^j$
and 
$\gbf_\Tbf = \lambda^2 \, dt^2 + g_{ij} \, dx^i dx^j$, respectively. 
By comparing the covariant derivative operators, $\Dbf$ and $\Dbf_{\gbf_\Tbf}$,  
of both metrics and relying on the Lie derivative bound (A2), we obtain  
$$ 
| \Dbf_{\gbf_\Tbf} - \Dbf|_{\gbf_\Tbf} \leq K_0 \, e^{K_0}. 
$$
Hence, using this estimate and computing the length of a Lorentzian geodesic $\gamma$ in terms 
of its Riemannian length, we control the radius of definition of the exponential map.   

Next, to control the radius of conjugacy associated with the exponential map
we estimate the length of Jacobi fields, that is, variations of geodesics defined as usual by  
$$
J : = {\del \over \del t} \gamma(s,t), 
\qquad 
\ddot{J} = - \Riem(\dot{\gamma}, \dot{\gamma}, J).
$$  
Here, the estimates use in an essential way, the curvature assumption (A3). 
Finally, we complete the proof by considering the radius of injectivity of the exponential map.
and showing that no two geodesic can intersect in a sufficiently small ball, at least.  
\end{proof}


Clearly, Theorem~\ref{premier} is not satisfactory since the notion of injectivity radius of 
an observer depends only upon the given vector $\Tbf_\pbf$ and not 
on the vector field $\Tbf$ which we introduced along with the foliation of a neighbhoorhood of the point. 
 This observation motivates the following discussion leading to the more general result in Theorem~\ref{deuxieme}.  
 
In fact, it is not necessary to prescribe a timelike vector field (or a foliation) a~priori 
and, instead, we can reconstruct geometrically and 
determine a ``canonical'' foliation adapted to the local geometry. 
So, the given data are now a single observer $ (\pbf,\Tbf_\pbf)$ which allows us to 
define the reference inner product $\gbf_{\Tbf_\pbf}$ at the point $\pbf$, only.   
We always assume $\expo_\pbf$ defined in $B_{\gbf_{\Tbf_\pbf}}(0,r)\subset T_\pbf\Mbf$ for some $r>0$. 

To state our main assumption that the curvature is bounded we need
a reference metric defined in a {\sl whole} neighborhood of $\pbf$. We proceed as
follows. By parallel transporting (with respect to the Lorentzian connection $\Dbf$) the given vector $\Tbf_\pbf$ 
along radial geodesics leaving from $\pbf$ we obtain a (possibly multivalued) vector field $\Tbf$ 
defined in a neighborhood of $\pbf$. In turn, we can define an inner product
$\gbf_\Tbf$ defined in the tangent space
$T_\qbf \Mbf$ for each $\qbf \in B_{\gbf_\Tbf}(\pbf, r)$ where the
exponential map is already well-defined. 
 
We define the {\sl maximum curvature for the observer} $(\pbf, \Tbf_\pbf)$ at the scale $r$
as  
$$ 
\Rmax(\Mbf,\gbf, \pbf,\Tbf_\pbf;r) := \sup_{\gamma} |\Riem_\gbf|_{\Tbf_\gamma}, 
$$ 
where the supremum is taken over all points along radial geodesics 
$\gamma: [0,r] \to \Mbf$ from $\pbf$ with length at most $r$.
 Note that when two distinct geodesics $\gamma$ and $\gamma'$ meet, 
$\Tbf_\gamma$ and $\Tbf_{\gamma'}$ are generally distinct.

The problem under consideration is equivalent to controling the geometry of the local covering 
$\expo_\pbf : B_{\gbf_{\Tbf_\pbf}}(0,r) \rightarrow \Bcal_{\gbf_{\Tbf_\pbf}}(\pbf,r) \subset \Mbf$.   

\begin{theorem}[Injectivity radius estimate for an observer]
\label{deuxieme} 
Let $(\Mbf,\gbf, \pbf, \Tbf_\pbf)$ be a pointed Lorentzian $(n+1)$-manifold
such that, for some scalar $r>0$,
the unit geodesic ball $\Bcal_{\gbf_{\Tbf_\pbf}}(\pbf,r)$ is compactly included in $\Mbf$ 
and the map 
$\expo_\pbf$ is defined on $B_{\gbf_\Tbf}(0,r)\subset T_\pbf\Mbf$ with 
$$ 
\Rmax(\Mbf,\gbf,\pbf,\Tbf_\pbf; r) \leq r^{-2}. 
$$
Then, for some $c(n) \in (0,1]$ depending on the dimension, only,  
one has 
$$ 
{\inj(\Mbf,\gbf,\pbf,\Tbf_\pbf) \over r} \geq c(n) \, \frac{\vol_\gbf(\Bcal_{\gbf_\Tbf}(\pbf,c(n) \, r))}{r^{n+1}}.
$$ 
\end{theorem}

\begin{proof} We only sketch the proof and, without loss of generality, take $r=1$. 
For the analysis, the vector field $\Tbf$ can not be used directly and, instead, 
it is necessary to construct a new vector field $\Nbf$.  
Fix $\qbf \in \Ical^-(\pbf)$ (the past of the point $\pbf$) 
and consider the ``time function'' $\tau := d_\gbf(\cdot, \qbf)$. 
The vector field $\Nbf := \nabla \tau$ is time-like and can be used as a (new) reference field
to which we associate the Riemannian metric $\gbf_\Nbf$. 
 
Recall that Hessians of distance functions are controlled by curvature, on which we precisely 
have a uniform bound. 
On the other hand, controling the Hessian $\nabla^2 \tau$ allows us to 
a control of the ``relative'' geometry of the slices and, in turn, the curvature of the reference metric 
$\gbf_\Nbf$.  
$$ 
|\Riem_{\gbf_\Nbf}|_{\gbf_\Nbf} \leq C. 
$$

At this stage, we are back to the situation studied in Theorem~\ref{premier}
and we can follow the same techniques and estimate the conjugate radius at $\pbf$.
The final estimate of the radius of injectivity of the exponential map, as stated in the theorem,
is more precise than what we derived earlier and
our final argument here is a Lorentzian generalization of 
an homotopy argument on geodesic loops due to Cheeger, Gromov, and Taylor \cite{CGT}
in the Riemannian setting. 
\end{proof}


\section{Local regularity of pointed Lorentzian manifolds}
\label{REGUL}

Given a pointed Lorentzian manifold $(\Mbf, \gbf, \pbf, \Tbf_\pbf)$ that solely 
satisfies curvature and injectivity radius bounds, 
our objective now is to establish the existence of a local coordinates chart 
defined in a ball with {\sl definite size,}
in which the metric coefficients have {\sl optimal regularity.}  
No further regularity of the metric beyond the curvature bound will be required. 

Our objective, now, is to construct a foliation around the
point $\pbf$,  
$$
\bigcup_{t \in [\bart(\pbf),\tbar(\pbf))} \Sigma_t,  
$$ 
by $n$-dimensional spacelike hypersurfaces $\Sigma_t \subset \Mbf$ with 
constant mean curvature $t$. 
The range of $t$ is specified by two functions $\bart(\pbf), \tbar(\pbf)$.

The novelty of the following theorem lies in the {\sl quantitative estimates} 
involving curvature and injectivity bounds, only.  
It  provides a {\sl canonical local foliation} for the given observer. 

\begin{theorem}[Local CMC foliation of an observer]
\label{folii}
The following property holds with constants $c, \underline \rho, \ldots \in (0,1)$ 
depending on the dimension $n$, only.
Let $(\Mbf, \gbf,\pbf,\Tbf_\pbf)$ be a pointed Lorentzian manifold satisfying,
at some scale $ r>0$,  
$$ 
\Rmax(\Mbf, \gbf, \pbf, \Tbf_\pbf; r) \leq r^{-2},
\qquad
\inj(\Mbf, \gbf, \pbf, \Tbf_\pbf) \geq r.
$$
Then, the Riemannian ball $ \Bcal_\Tbf(\pbf,cr)$ is covered by a foliation of 
spacelike hypersurfaces $ \Sigma_t$ with constant mean curvature $t \in [\bart(\pbf), \tbar(\pbf)]$
$$
\aligned 
& \Big( \bigcup_{\bart(\pbf) \leq t \leq \tbar(\pbf)} 
\Sigma_t \Big) \supset \Bcal_\Tbf(\pbf,cr),
\\
& r t \in \big[ (1-\eta) \rho, (1+\eta) \rho \big], 
\qquad 
\bart(\pbf) := {1-\zeta \over sr}, \quad \tbar(\pbf) := {1+\zeta \over sr} 
\endaligned
$$
for some $\rho \in [\underline \rho, \overline \rho]$ and $s \in [\barc,\cbar]$.  
Moreover, the unit normal $ \Nbf$, the lapse function $ \lambda^2 := -\gbf(\Dbf t, \Dbf t)$, 
and the second fundamental form $h$ of this foliation satisfy the uniform estimates 
$$
\aligned 
& 1 - \theta \leq  - \gbf(\Nbf,\Tbf) &   \leq 1,  
\qquad 
         && \theta \leq -r^{-2} \lambda\leq \theta^{-1},
\\
& r \, |h|_{\gbf_\Tbf}  \leq \theta^{-1}.             &&
\endaligned
$$
Recall that the vector field $\Tbf$ is defined by parallel translating the vector $\Tbf_\pbf$
along radial geodesics from $\pbf$. 
\end{theorem}

\

Our proof is a generalization of earlier work by Bartnik and Simon \cite{BartnikSimon}
(for hypersurfaces in Minkowski space) and Gerhardt \cite{Gerhardt} (providing global foliations of Lorentzian manifolds).

\begin{proof} We will only sketch the proof.  
First of all, we need  a {\sl Lorentzian geodesic foliation} near the observer $(\pbf,\Tbf_\pbf)$,  
that is, a foliation by geodesic spheres
$\bigcup_\tau \Hcal_\tau$. 
This foliation is constructed by considering 
the future-oriented, timelike geodesic $\gamma: [0, \cbar r] \to \Mbf$ such that 
$\gamma(c r) = \pbf$ and $\dot \gamma(\pbf) = \Tbf_\pbf$ for some (fixed once for all) constant $c \in (0,1)$, 
and by then introducing normal coordinates $y= (y^\alpha)=(\tau, y^j)$ based on  
radial geodesics from the point $\qbf := \gamma(0)$. 

Relying on our curvature bound together with a (Lorentzian) 
variant of the Hessian comparison argument theorem, we obtain   
$$ 
\bark(\tau, r) \, \gbf_{ij} \leq (-\Dbf^2\tau)|_{E, ij} \leq \kbar(\tau,r) \, \gbf_{ij},
$$
where $E:= \big( \Dbf \tau \big)^\perp$ denotes the orthogonal complement of the gradient
and for some $C>0$ 
$$ 
\bark(\tau,r) := \frac{r^{-1} \sqrt{C}}{\tan \big(\tau \, r^{-1} \sqrt{C}\big)},
\qquad
\kbar(\tau,r) := \frac{r^{-1} \sqrt{C}}{\tanh \big(\tau \, r^{-1} \sqrt{C}\big)}.
$$
Hence, by taking the trace of $(-\Dbf^2\tau)|_{E, ij}$ we can 
control the mean curvature $H_{\Hcal_\tau}$ of each geodesic slice:  
$$ 
n \, \bark(\tau,r) \leq H_{\Hcal_\tau}\leq n \, \kbar(\tau,r). 
$$

In a similar fashion, we can also construct a {\sl Riemannian geodesic foliation} near $\pbf$.
Roughly speaking,
we pick up a point $\pbf' = \gamma(\tau)$ in the future of $\pbf$
$\tau > c r$ 
and, then, for each $a$ within some interval  
we consider the Riemannian slice $\Acal(\pbf',a) := S_{\gbf_{\Tbf_\qbf}}(\pbf', a) \cap \Jcal^+(\qbf)$
determined by the reference metric $\gbf_{\Tbf_\qbf}$ associated with $\Tbf_\qbf$. 
Again, using the curvature bound and the standard Hessian comparison theorem 
we estimate the mean curvature of the Riemannian slices $H_{\Acal(\pbf',a)}$: 
$$ 
n \, \bark(a,r) \leq H_{\Acal(\pbf',a)} \leq n \, \kbar(a,r).
$$

We then search for the desired CMC foliation $\bigcup_t  \Sigma_t$ in such a way that 
each hypersurface $\Sigma_t= \big\{ (u^t(y),y) \big\}$ can be viewed 
as a graph over a (fixed) geodesic slice $\Hcal_\tau$. The heart of our construction 
lies in the derivation of uniform estimates for the mean curvature operator
$$ 
\Mcal u := h_{ij} g^{ij}
= {1 \over \sqrt{1 +|\nabla u|^2}}
\left(
\Delta u + {A_j}^j\right),  \qquad A := \nabla^2 \tau. 
$$
The Lorentzian and Riemannian slices above played the role of barrier functions for this operator.

To precisely ``localize'' a CMC slice, we fix a real $s$ in some interval $[\barc,\cbar]$ and 
we consider the point $\pbf_s := \gamma((s+s^2)r)$. Then, we consider the domain 
$\Omega_s \subset \big\{ \tau=s r \big\}$ defined so that its boundary is 
$$  
\del \Omega_s := \Acal\big( \pbf_s, (s^2+s^3)r \big) \cap \{\tau=sr\}, 
$$
which implies that  
$$ 
B_{sr}\big( \gamma(sr), s^{5/2}r/2 \big) \subset \Omega_s \subset B_{sr}\big( \gamma(sr), 2s^{5/2} r \big).
$$ 
The CMC slice is searched as a graph over $\Omega_s$ and, 
in particular, we establish via a gradient estimate that this slice is uniformly spacelike, 
as required. 
Our analysis uses the so-called Simons identity satisfied by the second fundamental form 
and arguments from the Nash-Moser technique. 
\end{proof} 
 
 
Our final objective is to introduce suitable coordinates in which the coefficient 
of the Lorentzian metric have the best possible regularity, under the sole regularity assumptions
that the curvature is bounded. It turns out that {\sl CMC-harmonic coordinates} 
provides the best choice; our result below covers 
manifolds satisfying vacuum Einstein equations, that is, 
the class of Ricci-flat Lorentzian manifolds.  

\begin{theorem}[Canonical cordinates of an observer]
\label{canon} 
There exist constants $\barc=\barc(n) < c=c(n) $ and $C=C(n,q) >0$ 
depending upon the dimension $n$ 
(and some exponent $q \in [1,\infty))$ such that 
the following properties hold.
Let $(\Mbf,\gbf, \pbf,\Tbf_\pbf)$ be an $(n+1)$-dimensional, 
pointed  vacuum  Einstein spacetime satisfying the 
following curvature and injectivity bounds at the scale $r>0$: 
$$
\Rmax(\Mbf, \gbf, \pbf, \Tbf_\pbf; r) \leq r^{-2},
\qquad
\inj(\Mbf, \gbf, \pbf, \Tbf_\pbf) \geq r.
$$
Then, there exist local coordinates $ \xbf=(t,x^1,\ldots,x^n)$ defined for all
$$ 
|t-r_1| < c^2 r, \qquad \big( (x^1)^2 + \ldots + (x^n)^2 \big)^{1/2} < c^2 r,
$$
such that $x(\pbf) = {(r_1,0,\ldots,0)}$ for some $r_1\in [\barc r, c r]$
and the following properties hold. 

Each hypersurface $\Sigma_t = \big\{(x^1)^2 + \ldots + (x^n)^2 < c^4 r^2\big\}$ 
is a spacelike hypersurface with constant mean curvature $c^{-1} r^{-2} t$.
The coordinates $x:=(x^1,\ldots,x^n)$ are spatially harmonic for Riemannian metric induced on $\Sigma_t$.
Moreover, in the coordinates $\xbf=(t,x^1,\ldots,x^n)$ 
the Lorentzian metric reads 
$$ 
\gbf = - \lambda(\xbf)^2 \, (dt)^2 + g_{ij} (\xbf) \big( dx^i + \xi^i(\xbf) \, dt \big) \big( dx^j + \xi^j(\xbf) \, dt \big)
$$
and is close to Minkowski metric in these local coordinates, in the sense that
$$ 
\aligned 
& e^{-C} \leq \lambda \leq e^C,
\qquad  
e^{-C}\delta_{ij}\leq g_{ij} \leq  e^{C}\delta_{ij},
\\
& 
|\xi|_\gbf^2 := g_{ij}\xi^i \xi^j \leq e^{-C}, 
\endaligned
$$
and for all $q \in [1,\infty)$ and for some $Q(n,q)>0$ 
$$ 
r^{-n+q} \int_{\Sigma_t}|\del \gbf|^q \, dv_{\Sigma_t}
+
r^{-n+2q} \int_{\Sigma_t}|\del^2 \gbf|^qdv_{\Sigma_t} \leq Q(n,q).
$$ 
\end{theorem}

  
\section{Injectivity radius of null cones}

The theory in Section~\ref{INJECT} can be extended to null cones. Given an observer $(\pbf,\Tbf_\pbf)$
in a Lorentzian manifold, consider 
its past cones in both the tangent space at $\pbf$ and the manifold, defined by  
$$ 
\aligned 
 N_\pbf^- &  := \big\{ X \in T_\pbf\Mbf \, \big/ \, \gbf_\pbf(X,X) = 0, \, \gbf_\pbf(\Tbf_\pbf,X) \geq 0 \big\}
\endaligned 
$$
and ${ \Ncal^-(\pbf)}  := \del \Jcal^-(\pbf)$.
Given $r>0$, the restriction of the exponential map to the the cone is called
the {\sl null exponential map} and is denoted by
 $\expo^N_\pbf : B_{\gbf_\Tbf,\pbf}^N(0,r) \to \Ncal^-(\pbf)$, 
where 
$B_{\gbf_\Tbf,\pbf}^N(0,r) := B_{\gbf_\Tbf,\pbf}(0,r)\cap N_\pbf^-$.

\begin{definition} The {\sl past null injectivity radius of an observer} $(\pbf,\Tbf_\pbf)$, 
$$
\NullInj^-(\Mbf,\gbf,\pbf,\Tbf_\pbf),
$$
is the supremum among all radii $r$ such that $ \expo_\pbf^N$ is a global diffeomorphism 
from $B_{\gbf_\Tbf,\pbf}^N(0,r) \backslash\{0\}$ to a pointed neighborhood of $\pbf \in \Ncal^-(\pbf)$.
\end{definition}

We make the following assumptions on a foliation $\Omega = \bigcup_{t \in [-1,0]} \Hcal_t$, 
with  unit normal $\Tbf$ and lapse function $ \lambda$, normalized so that $\pbf \in \Hcal_0$:  
\begin{enumerate}

\item[] $(A1) \quad   e^{-K_0} \leq \lambda \leq e^{K_0}$ in $\Omega$. 

\item[] $(A2) \quad   \sup_\Omega |\Lie_\Tbf \gbf |_{\gbf_\Tbf} \leq K_0$. 

\item[] $(A3') \, \text{ The null conjugate radius at }  p$ is $ \geq  r$ and 
in $ B^N := B_{\gbf_\Tbf,\pbf}^N(0,r)$ the null exponential map satisfies
$$ 
e^{-K_0} \, \gbf_{\Tbf,\pbf} \mid_{B^N} \leq
\big( {\expo^N_\pbf}\big)^\star \big( \gbf_T \mid_{\Bcal^N} \big)
\leq
e^{K_0} \, \gbf_{\Tbf,\pbf}\mid_{B^N}. 
$$

\item[] $(A4')$ \, There exist coordinates on $ \Hcal_{-1}$ such that 
$g\mid_{\Hcal_{-1}}$ is comparable to the Euclidian metric   
$$ 
e^{-K_0}\, \gbf_{E'} \leq g\mid_{\Hcal_{-1}} \leq e^{K_0} \, \gbf_{E'}
\quad 
\text{ in } \Bcal_{\Hcal_{-1}, E'}(\pbf,r).  
$$ 
The latter condition follows, for instance, from curvature and volume bounds on the 
``initial'' hypersurface $\Hcal_{-1}$. 
\end{enumerate}

\begin{theorem}[Null injectivity radius estimate]
\label{null3}
Let $(\Mbf,\gbf,\pbf,\Tbf_\pbf)$ be a pointed Lorentzian manifold satisfying the regularity assumptions 
$(A1)$, $(A2)$, $(A3')$, and $(A4')$.
Then, there exists a positive constant $I_0=I_0(K_0,r,n)$ such that 
$$ 
\NullInj^-(\Mbf,\gbf,\pbf,\Tbf_\pbf) \geq I_0.
$$
\end{theorem}

\begin{proof} First, we construct coordinates $(x^\alpha) = (t, x^j)$ near $\pbf$
in which the metric $ \gbf_T$ is comparable with the Euclidian metric $ \gbf_E$ in these coordinates. 
Then, we establish uniform estimates that ``localize'' the null cone within the region limited by
 two ``flat'' null cones: 
$$ 
\Ncal^-(\pbf) \cap \Hcal_t\subset \Acal^t_{[c_1 \, |t|, C_1|t|]},
\qquad t\in [-c_1 \, r,0], 
$$ 
$$ 
\Acal^t_{[a,b]} := \big\{ x^0 = t, \quad a^2 < (x^1)^2 + \ldots + (x^n)^2 < b^2 \big\} \subset \Hcal_a. 
$$
We obtain a Lipschitz continuous parametrization of the null cone and, in turn, 
we can estimate the injectivity radius from an homotopy argument restricted to the null cone. 
\end{proof}

We can combine the result above with an earlier theorem by Klainerman and Rodnianski 
on the conjugacy radius of null cones (\cite{KR5} and the references therein): 
when $ n=3$ and the manifold satisfies the  vacuum Einstein equations (Ricci-flat condition), 
Assumption $(A3')$ is a consequence of the following $L^2$ curvature bound on the initial hypersurface: 
\begin{enumerate}

\item[] $(A3'') \quad \|\Riem_\gbf\|_{\Lbf^2(\Hcal_{-1},\gbf_T)} \leq K_0$.

\end{enumerate} 
Hence, from Theorem~\ref{null3} we can deduce that the null injectivity radius of an observer in a
 vacuum  Einstein spacetime
is uniformly controled solely in terms of the
 lapse function, the second fundamental form of the foliation, and 
 the $L^2$ curvature and lower volume bounds on some initial hypersurface.

\begin{remark}
The condition $(A3')$ is a weaker version of $(A3)$ and we expect that it should hold when the curvature 
in every spacelike hypersurface is bounded in $L^{{n \over 2} + \eps}$ for some $\eps>0$. 
On the other hand, the condition $(A4')$ on the initial hypersurface is only ``slightly'' stronger 
than the volume bound $(A4)$ assumed earlier.  
\end{remark}


\section{Concluding remarks}

We conclude this text with possible extensions of the present work. 
In the context of Riemannian geometry, Anderson and Petersen (see \cite{Petersen} for a review)
have introduced a notion of {\sl harmonic radius} for Riemannian manifolds and 
established pre-compactness results for sequences of manifolds whose harmonic radius is uniformly bounded below.  
Similarly, based on our results for Lorentzian manifolds, it should be possible to define a notion of 
{\sl CMC-harmonic radius} ${\mathbf r}_{a,Q}(\Mbf,\gbf, \pbf, \Tbf_\pbf)$ and
 establish corresponding pre-compactness theorems for sequences of pointed Lorentzian manifolds 
whose CMC--harmonic radius is bounded below. Given reals $a>1$ and $Q,r >0$,
we expect that the class of $(n+1)$-dimensional, pointed Lorentzian manifolds
$$ 
{\mathcal E}_n^{a,Q}(r) := \left\{ (\Mbf,\gbf,\pbf, \Tbf_\pbf) \, : \,
{\mathbf r}_{a,Q}(\Mbf,\gbf, \pbf, \Tbf_\pbf) \geq r \right\}
$$
is strongly pre-compact in $ W^{l,a}$ for $l \in [0,2)$ and 
weakly pre-compact in $ W^{2,a}$.

In fact, our main result (Theorem~\ref{canon}) should be restated as 
a uniform lower estimate on the CMC-harmonic radius, under curvature and injectivity radius bounds. 
In turn, by combining the above two statements, one arrives at a pre-compactness theorem for 
sequences of vacuum spacetimes with uniformly bounded curvature and injectivity radius bounded below. 

In another direction, we expect the techniques in this paper to be useful in constructing 
a canonical CMC foliation near spacelike infinity $i^0$, again with a control of the geometry that 
only depends on the sup-norm of the curvature. 
 
In conclusion, our results show that a bound on the curvature allows one
to get optimal control on the geometry of pointed Lorentzian manifolds. 
In contrast with Riemannian geometry where harmonic coordinates provide the best regularity of metrics, 
in the Lorentzian setting one needs a foliation by Constant Mean Curvature slices 
and spatially harmonic coordinates.
Another particular feature of Lorentzian geometry is the need of choosing an observer 
on the spacetime. 
Provided with the key regularity properties in
Theorems~\ref{folii} and \ref{canon}, we have 
now the necessary tool to tackle questions
about convergence and compactness of spacetimes. A long-term goal of this research will be
to analyze the structure of 
the future boundary of a spacetime (nature of singularities, curvature blow-up, relation with 
Penrose conjecture). 
 

\section*{Acknowledgements}

I am very grateful to G\'erard Besson for providing me the opportunity to present this work in the 
``S\'eminaire de Th\'eorie Spectrale et de G\'eom\'etrie'' at the Universit\'e de Grenoble and to have fruitful discussions
with the members of the Institut Fourier. 
This work was partially supported by the Agence Nationale de la Recherche (ANR) through the
grant 06-2-134423 entitled {\sl ``Mathematical Methods in General Relativity''} (MATH-GR).

Part of this work was written when the author was visiting the Institut Henri Poincar\'e
in the Spring 2008
during the Semester Program ``Ricci Curvature and Ricci Flow''  
organized by G. Besson, J. Lott, and G. Tian.
This paper was completed when the author visited the Mittag-Leffler Institute
in the Fall 2008 during the Semester Program ``Geometry, Analysis, and General Relativity''
organized by L. Andersson, P. Chrusciel, H. Ringstr\"om, and R. Schoen.


\end{document}